\documentclass[fleqn,usenatbib]{mnras}

\usepackage{newtxtext,newtxmath}

\usepackage[T1]{fontenc}

\DeclareRobustCommand{\VAN}[3]{#2}
\let\VANthebibliography\thebibliography
\def\thebibliography{\DeclareRobustCommand{\VAN}[3]{##3}\VANthebibliography}

\usepackage{graphicx}	
\usepackage{amsmath}	
\usepackage{listings}
\usepackage{xcolor}
\usepackage{color, colortbl}

\newcommand\crule[3][black]{\textcolor{#1}{\rule{#2}{#3}}}
\definecolor{C0}{HTML}{4c72b0}
\definecolor{C1}{HTML}{dd8452}
\definecolor{C2}{HTML}{55a868}
\definecolor{C3}{HTML}{c44e52}
\definecolor{C4}{HTML}{8172b3}
\definecolor{C6}{HTML}{da8bc3}
\definecolor{C8}{HTML}{ccb974}
\definecolor{C9}{HTML}{64b5cd}
\usepackage{multirow}
\usepackage{xcolor,pifont}
\newcommand{\cmark}{\textcolor{green!80!black}{\ding{51}}}
\newcommand{\xmark}{\textcolor{red}{\ding{55}}}

\title[]{New constraint on the tensor-to-scalar ratio from the \textit{Planck} and BICEP/Keck Array data using the profile likelihood}

\author[Paolo Campeti and Eiichiro  Komatsu]{
Paolo Campeti $^{1,2}$\thanks{E-mail: pcampeti@mpa-garching.mpg.de} and Eiichiro  Komatsu$^{1,3}$
\\
$^{1}$Max Planck Institute for Astrophysics, Karl-Schwarzschild-Str.1, 85741 Garching, Germany\\
$^{2}$Excellence Cluster ORIGINS, Boltzmannstr. 2, 85748 Garching, Germany\\
$^{3}$Kavli Institute for the Physics and Mathematics of the Universe (Kavli IPMU, WPI),
UTIAS, The University of Tokyo, Chiba, 277-8583, Japan
}

\date{Accepted XXX. Received YYY; in original form ZZZ}

\pubyear{2022}

\begin{document}
\label{firstpage}
\pagerange{\pageref{firstpage}--\pageref{lastpage}}
\maketitle

\begin{abstract}
We derive a new upper bound on the tensor-to-scalar ratio parameter $r$ using the frequentist profile likelihood method. We vary all the relevant cosmological parameters of the $\Lambda$CDM model, as well as the nuisance parameters. Unlike the Bayesian analysis using Markov Chain Monte Carlo (MCMC), our analysis is independent of the choice of priors. Using \textit{Planck} Public Release 4, BICEP/Keck Array 2018, \textit{Planck} CMB lensing, and BAO data, we find an upper limit of $r<0.037$ at 95\% C.L., similar to the Bayesian MCMC result of $r<0.038$ for a flat prior on $r$ and a conditioned \textit{Planck} lowlEB covariance matrix. 
\end{abstract}

\begin{keywords}
cosmology: cosmic background radiation -- methods: data analysis -- cosmology: cosmological parameters
\end{keywords}



\section{Introduction}\label{sec:intro}
Detecting the stochastic background of primordial gravitational waves predicted within the inflationary paradigm \citep{Grishchuk:1974ny, Starobinsky:1979ty} represents one of the principal objectives of the current cosmological research, as it would provide the definitive evidence for cosmic inflation \citep{Guth:1980zm, Sato:1980yn, Linde:1981mu, Albrecht:1982wi}.

Whereas inflation produces gravitational waves (i.e. tensor modes) over a wide range in frequency measurable by several different probes \citep[see e.g.][for a review]{Campeti:2020xwn}, the most promising route to detection is the $B$-mode polarization of the cosmic microwave background (CMB) \citep{Kamionkowski:1996zd, Seljak:1996gy}.

The current datasets only provide upper bounds on the tensor-to-scalar ratio $r$ (i.e. the ratio of the amplitudes of the tensor and scalar modes power spectra).
To date, the tightest limit on $r$ (customarily measured at the pivot scale $k_0 = 0.05$ ${\rm Mpc^{-1}}$) is $r<0.032$ at $95\%$ C.L. \citep{Tristram:2021tvh}, coming from the \textit{Planck} latest CMB temperature and $E$ and $B$-mode polarization data \citep{Tristram:2020wbi}, the BICEP/Keck Array $B$-mode data \citep[][hereafter BK18]{BICEP_2021}, the baryon acoustic oscillations (BAO) of the large-scale structure \citep{eBOSS:2020yzd}, and the CMB lensing data \citep{Planck:2018lbu}. This upper limit is derived using a standard Bayesian Monte Carlo Markov Chain (MCMC) procedure, varying the relevant cosmological parameters of a 
flat $\Lambda$ cold dark matter ($\Lambda$CDM) model and adopting the \citet{Sellentin:2015waz} correction (hereafter SH) to the \citet{Hamimeche:2008ai} likelihood (hereafter HL) for the \textit{Planck} large-scale $EE$, $BB$ and $EB$ power spectra (the ``lowlEB'' likelihood). The SH correction is needed to account for the increased uncertainty in parameter estimation due to the limited number of simulations used to estimate the covariance matrix. This is obtained by analytically marginalizing over the unknown true covariance matrix.

Most of the constraining power on $r$ at the pivot scale comes from BK18's $B$-mode data. An upper limit of $r<0.036$ at $95\%$ C.L. \citep{BICEP_2021} is obtained just from the BK18 data, provided that we fix the $\Lambda$CDM parameters to their best-fitting values given in \citet{planck_2018}. 
The \textit{Planck} satellite provides, on the other hand, the tightest constraints to date on the $B$ modes at the largest angular scales, which are not accessible from the ground. Exploiting the latest \texttt{NPIPE}-processed Public Release 4 (PR4) of temperature and polarization maps, the \textit{Planck} collaboration reported a limit of $r<0.056$ at $95\%$ C.L. \citep{Tristram:2020wbi}, which is relaxed to $r<0.075$ when properly accounting for the SH correction in the lowlEB likelihood \citep{Beck:2022efr}.

While the SH correction accounts for the Monte Carlo noise in the estimated covariance matrix, it does not correct for the additional scatter in the best-fitting maximum a posteriori parameter (MAP) estimate, which can lead to a misestimation of confidence limits \citep{Beck:2022efr}. This effect is especially relevant near the physical boundary of a given parameter (i.e. $r\geq 0$ in our case of interest) and can produce a significant underestimation of the upper limit. The issue can be corrected by increasing the number of (computationally expensive) time-ordered data simulations used in the covariance matrix estimation or by properly conditioning the covariance matrix. The latter method has been applied in \citet{Beck:2022efr} to the lowlEB \textit{Planck} likelihood (which we will refer to as ``conditioned HL'' in the following), resulting in a much weaker \textit{Planck}-only upper limit of $r<0.13$ at 95\% C.L., associated to a large shift of the peak of the marginalized distribution to larger $r$ values than in the SH case. Similarly, for the \textit{Planck} + BK18 + BAO + lensing combination, the conditioning results in a more conservative upper limit of $r<0.038$. 

In this paper, we present constraints on $r$ using the frequentist profile likelihood method, and compare them to the standard Bayesian MCMC procedure adopted throughout the literature\footnote{We emphasize that all upper limits on $r$ reported above have been derived with an MCMC approach.}. While the profile likelihood is a standard data analysis tool in particle physics \citep[see e.g.][]{ParticleDataGroup:2020ssz, ATLAS:2013mma}, it has been seldom used in cosmology, notable cases of use being the application to $\Lambda$CDM parameters estimation from the \textit{Planck} data \citep{planck_profile}, to the Early Dark Energy fraction \citep{Herold:2021ksg}, to coupled dark energy and Brans-Dicke models \citep{Gomez-Valent:2022hkb} and to the estimation of $r$ from the SPIDER data \citep{SPIDER:2021ncy}.  Nonetheless, this approach bears several potentially interesting differences with Bayesian methods \citep{Cousins:1994yw}. First, the profile likelihood does not require priors, which may have an impact on the final constraints. Second, while in Bayesian methods the choice of a specific set of parameters to sample might represent an implicit prior choice, the maximum likelihood estimate (MLE) is invariant under model reparameterization. Third, the parameter estimates obtained from the profile likelihood are not affected by ``volume effects'' which can arise during marginalization in the MCMC approach \citep{Hamann:2007pi}. Moreover, the profile likelihood formalism allows to conveniently include the effect of the parameter's physical boundary in the confidence intervals via the Feldman-Cousins prescription \citep{Feldman:1997qc}.   

Our work aims to deconstruct the current constraints on $r$ and scrutinise their robustness. Similarly to the profile likelihood analysis performed on the $\Lambda$CDM parameters \citep{planck_profile}, we study the effect of priors and marginalization on  the inference of $r$ from the \textit{Planck} and BK18 data. We also explore the effect of conditioning the \textit{Planck} lowlEB covariance matrix \citep{Beck:2022efr}, on the profile likelihood.

The structure of the paper is the following. We describe the data and likelihood used in our analysis in Section \ref{sec:data}. We review the profile likelihood formalism and the Feldman-Cousins prescription in Section \ref{sec:profile}. We discuss the new constraints on $r$ from our frequentist analysis and compare them to the Bayesian credible intervals in Section \ref{sec:results}. We conclude in Section \ref{sec:conclusions}.

\section{Data and likelihoods}\label{sec:data}

We use the latest \textit{Planck} \texttt{NPIPE}-processed PR4 maps \citep{Tristram:2020wbi} and the BK18 dataset \citep{BICEP_2021}.
We use the data and likelihoods publicly available for the $\texttt{Cobaya}$\footnote{\url{cobaya.readthedocs.io}} \citep{Torrado:2020dgo} MCMC framework, as done in \citet{Tristram:2021tvh}. We also use \texttt{Cobaya} as an interface with the \texttt{CAMB} Boltzmann solver \citep{camb}. 

\subsection{\textit{Planck} likelihoods}
The \textit{Planck} likelihood consists of three parts: the low-$\ell$ $TT$ \texttt{Commander} likelihood \citep{planck_2018_like} for $\ell=2-30$, the high-$\ell$ $TT+TE+EE$ \texttt{HiLLiPoP} likelihood\footnote{\url{github.com/planck-npipe/hillipop}}  \citep{Planck:2013win, planck_2015_like, Couchot:2016vaq} for $\ell = 30 - 2500$, and the low-$\ell$ $EE+BB+EB$ \texttt{LoLLiPoP} or the lowlEB likelihood\footnote{\url{github.com/planck-npipe/lollipop}} \citep{Tristram:2020wbi} for $\ell = 2 -150$. 

The low-$\ell$ $TT$ likelihood is the same as in PR3, since no improvement is expected with the PR4 update for the high signal-to-noise temperature data. The \texttt{HiLLiPoP} likelihood is instead a Gaussian likelihood for cross-power spectra of the \textit{Planck} 100, 143, and 217-GHz data. 

The \texttt{LoLLiPoP} likelihood for large-scale $EE$, $BB$ and $EB$ power spectra implements the HL  approximation for a non-Gaussian likelihood \citep{Hamimeche:2008ai}, adapted specifically for cross-power spectra \citep{Mangilli:2015xya}. In this case, an offset term is needed to make the distribution of cross-power spectra similar to that of auto-power spectra, as required by the HL approximation. The covariance matrix for this likelihood is estimated from 400 Monte Carlo simulations of PR4, which include \textit{Planck} noise, systematic effects and foreground residuals. The \textit{Planck} lowlEB likelihood implements the SH correction to the HL likelihood to account for the Monte Carlo noise in the covariance matrix estimate. This is not sufficient to amend the additional scatter in the MAP estimate: a possible solution indicated in \citet{Beck:2022efr} involves using the HL likelihood (without the SH correction) with a conditioned covariance matrix.
The conditioning strategy removes all off-diagonal elements beyond the next-to-nearest neighbour for unbinned multipoles ($\ell\leq 35$) and all off-diagonal elements beyond the nearest neighbour for binned multipoles  ($\ell> 35$). We will refer to this specific choice as ``cond. HL'' in the following.

\subsection{BICEP/Keck Array 2018 likelihood}
The BK18 likelihood, which includes only $B$ modes at $\ell \simeq 30 -300$, also applies the HL approximation to auto- and cross-power spectra in conjunction with the \textit{WMAP} data at 23 and 33 GHz and \textit{Planck} \texttt{NPIPE}-processed data at 30, 44, 143, 217 an 353 GHZ. The bandpower covariance matrix is estimated from 499 simulations. The default BK18 likelihood already incorporates conditioning to reduce the Monte Carlo noise. 

\subsection{Likelihood combination and priors in the default analysis}\label{sec:priors}
We combine the \textit{Planck} and BK18 likelihoods neglecting correlations between them. This is a good approximation because the current $B$-mode data are noise-dominated, the two CMB surveys have uncorrelated noises, and they observe very different fractions of the sky \citep[i.e. 50\% for \textit{Planck} and 1\% for BK18, see][]{Tristram:2021tvh, Tristram:2020wbi}. 
In the following, whenever we use the \textit{Planck} likelihood, we will also include the BAO data \citep{eBOSS:2020yzd} and the \textit{Planck} CMB lensing data \citep{Planck:2018lbu}. 

There are in total 33 free parameters in the default \textit{Planck} + BK18 analysis, including $r$, 6 parameters of a flat $\Lambda$CDM model $\{\Omega_{\rm b}h^2, \Omega_{\rm  c}h^2, \tau, A_{\rm  s}, n_{\rm  s}, \theta_{\rm MC}\}$, and the nuisance parameters. The tensor spectral index $n_{\rm t}$ is fixed via the inflationary consistency relation $n_{\rm t}=-r/8$, similarly to previous analyses \citep{Tristram:2021tvh, Tristram:2020wbi}. We also checked that fixing $n_t=0$ as in the \citet{BICEP_2021} analysis does not impact our results. 

The \textit{Planck} likelihoods introduce 19 nuisance parameters, accounting for map and absolute calibration and foreground modeling \citep[for a description see Appendix B in][]{Tristram:2020wbi}. Of these, 8 parameters have a Gaussian prior in the default MCMC analysis, whereas the others have uniform priors.
The BK18 likelihood has 7 nuisance parameters accounting for Galactic dust and synchrotron foreground modeling. Of these, 6 parameters have uniform priors in the default BK18 analysis \citep{BICEP_2021}, whereas the synchrotron spectral index $\beta_{\rm s}$ has a Gaussian prior $\beta_{\rm s}=-3.1\pm 0.3$ \citep[motivated by the WMAP 23 and 33 GHz data,][]{Fuskeland:2014eoa}.
As shown in \citet{BICEP_2021}, the constraint on $\beta_{\rm s}$ from the BK18 data is prior-dominated; therefore, for a more direct comparison with the Bayesian results in the literature, we also explore the possibility of fixing $\beta_{\rm s}=-3.1$ in the profile likelihood, since frequentist analyses do not incorporate priors. We indicate such choice as ``fixed $\beta_{\rm s}$'' in the following.

\section{Profile likelihood}\label{sec:profile}

We use the profile likelihood to investigate the effects of priors and marginalization on the current Bayesian constraints on $r$. The profile likelihood is a staple in the frequentist's toolbox. As it does not incorporate priors, explicitly or implicitly via the model parametrization, it is immune to volume effects which may appear during marginalization in MCMC.

The profile likelihood for a parameter of interest $\mu$ (in our case $\mu= r$) is obtained by fixing $\mu$ to multiple values within the range of interest and minimising the $\chi^2(\mu)=-2\log\mathcal{L}(\mu)$ with respect to all the remaining cosmological and nuisance parameters for each fixed value of $\mu$. Here,  $\mathcal{L}$ is the likelihood. By construction the minimum $\chi^2_{\rm min}$ coincides with the global MLE (also called ``best-fit''). 

We use $\Delta\chi^2(\mu)=\chi^2(\mu) - \chi^2_{\rm min}$ to construct frequentist confidence intervals on $\mu$. If $\mu$ is far away from its physical boundary, a confidence interval at $\alpha$ C.L. can be obtained by cutting $\Delta\chi^2(\mu)$ at a fixed threshold $\Delta\chi^2_{\rm th}$ such that the cumulative distribution function of the $\chi^2$ distribution with one degree of freedom is equal to $\alpha$ \citep[e.g. cutting at $\Delta\chi^2_{\rm th}=1$ and $\Delta\chi^2_{\rm th}=3.84$ for 68\% or 95\% C.L., respectively, see e.g.][]{Trotta:2017wnx}. We can use this procedure for both parabolic (associated to a Gaussian-distributed parameter) and nonparabolic $\Delta\chi^2(\mu)$ thanks to  invariance of the MLE under reparametrization.

\subsection{The Feldman-Cousins prescription}
If the parameter estimate is instead close to its physical boundary, as in our case of interest, the classical Neyman's construction of frequentist confidence intervals is unsatisfactory. It can lead to empty intervals and to failure of the frequentist coverage property\footnote{The frequentist coverage property is realized at the level $\alpha$ if a fraction $\alpha$ of the confidence intervals obtained from Neyman's construction contains the fixed and unknown true value of the parameter of interest.  \citep[see e.g.][]{Cousins:2018tiz}.} if the choice of reporting an upper limit or a two-sided interval is made by looking at the data.

These issues can be solved by adopting the \citet{Feldman:1997qc} (hereafter FC) prescription. For each value $\mu$ of the parameter of interest (with unknown true value) and each observable $x$, we compute the \textit{likelihood ratio}
\begin{equation}\label{eq:lkl_ratio}
    R(x, \mu) = \frac{\mathcal{L}(x|\mu)}{\mathcal{L}(x|\mu_{\rm best})},
\end{equation}
where $\mu$ can take only physically allowed values and $\mu_{\rm best}$ is the value of $\mu$ which maximizes the likelihood $\mathcal{L}(x|\mu)$. 
The so-called \textit{confidence belt} at the desired $\alpha$ C.L. is then built by selecting for each $\mu$ an \textit{acceptance interval} $[x_1, x_2]$ such that
\begin{equation}\label{eq:system}
    \begin{cases}
     R(x_1, \mu) = R(x_2, \mu), \\
      \int_{x_1}^{x_2} P(x|\mu) dx =\alpha,
    \end{cases}
\end{equation}
where $P(x|\mu)$ is the probability density function (pdf) for $x$ given $\mu$ and the values $x$ are added to the acceptance interval in order of decreasing likelihood ratio. The confidence belt is then given by the union of all acceptance intervals $[x_1(\mu), x_2(\mu)]$: intercepting it with a line at $x=x_0$, with $x_0$ being the value of $x$ minimising $\chi^2$ (i.e. the value measured in the experiment), we obtain the confidence interval $[\mu_1, \mu_2]$ for the parameter $\mu$.  

The FC prescription provides, therefore, an additional criterion to fix the extrema of the confidence intervals in Neyman's construction and to transition between an upper limit and a two-sided interval, giving exact frequentist coverage for a Gaussian parameter even in proximity of a physical boundary. This is in contrast with the conservatism (i.e. overcoverage) inherent to Bayesian limits in the same context \citep{Cousins:1994yw, Feldman:1997qc}. While conservatism might not be as a severe issue as undercoverage, it certainly degrades our ability to discriminate against false hypotheses, making it worthwhile to examine frequentist intervals. 

As we will see in the next section, the profile likelihood for $r$ gives a parabolic $\Delta\chi^2$ near its minimum, and presents a physical boundary at $r=0$. In this case, $\mu_{\rm best}=\max(0,\,x)$ and the likelihood ratio in Eq.\ref{eq:lkl_ratio} becomes \citep{Feldman:1997qc}:
\begin{equation}
    R(x, \mu) =  \begin{cases} 
    \exp(-(x-\mu)^2/2),\,&\text{for}\, x\geq0,\\
    \exp(x\mu-\mu^2/2),\,&\text{for}\, x < 0,
\end{cases}
\end{equation}
where $x$ and $\mu$ are expressed in units of $\sigma$, that is, the width of the parabolic fit to $\Delta\chi^2(\mu)$. The confidence interval is obtained by solving the system given in Eq.\ref{eq:system} with $P(x|\mu)$ being a Gaussian with mean $\mu$ and unit variance.

\subsection{Minimization algorithm}
We minimise $\chi^2(r)$ with the \texttt{MIGRAD} algorithm implemented in the  \texttt{iMinuit}\footnote{\url{iminuit.readthedocs.io}} package, a \texttt{python} interface for the popular \texttt{Minuit} multi-dimensional minimiser\footnote{We found significantly better performances using \texttt{iMinuit} compared to other common minimising algorithms (e.g. the \texttt{scipy} \texttt{minimize} module \citep{scipy} and \texttt{Py-BOBYQA}  (\url{numericalalgorithmsgroup.github.io}) which are already implemented in the \texttt{Cobaya} sampler).}.
We scan the parameter space, fixing $r$ to values over a wide range and minimising $\chi^2(r)$  with respect to the remaining 32 free parameters
for each fixed $r$. Each point in the profile likelihood typically requires around $\mathcal{O}(10^{4})$ evaluations of the likelihood, with each evaluation taking $\mathcal{O}(1)$ s (using 10 logical CPUs on a computer cluster node), almost exclusively absorbed by the evaluation of the \texttt{CAMB} Boltzmann code\footnote{The profile likelihood is highly competitive with the more traditional MCMC approach, which requires $\mathcal{O}(10^6)$ points to reach convergence, due to inefficient sampling of the Metropolis-Hastings algorithm near the boundary of a parameter with a uniform positive prior.}. 
To increase the chance of the minimiser reaching the global minimum, each minimization is started from ten different random initial parameter sets.
We take the point with the lowest $\chi^2$ as the final result. We checked that increasing the accuracy settings of the \texttt{CAMB} code does not change our results.

\section{Results and comparison to the Bayesian analysis}\label{sec:results}

   \begin{figure*}
    \centering
    \includegraphics[scale=0.7]{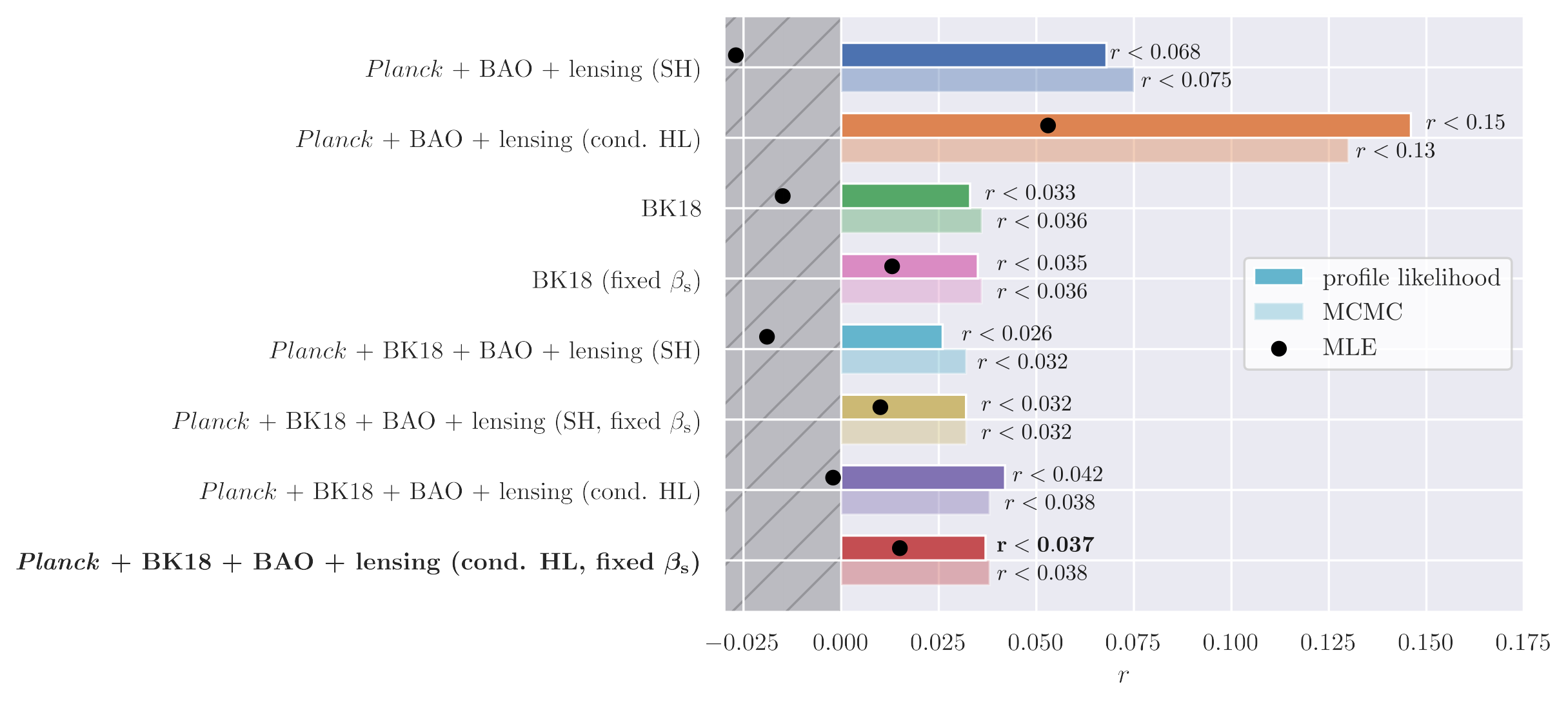}
    \caption{Summary of 95\% C.L. upper limits on $r$ for datasets considered in this work. The darker shaded bars indicate the upper limit from the profile likelihood, whereas the lighter shaded bars the MCMC one. The best-fitting $r$ values from the profile likelihood analysis are shown as the black dots. Negative (unphysical) values of $r$ are indicated by the hatched dark grey region. The baseline result of this work is highlighted in bold.}
    \label{fig:barplot}
\end{figure*}

In Fig. \ref{fig:barplot} and Table \ref{table:tab1} we report 95\% C.L. upper limits on $r$ obtained from the profile likelihood (darker shaded bars) and compare them to their Bayesian MCMC counterparts (lighter shaded bars). We also show the best-fitting $r$, that is, the global MLE values found in the profile likelihood analysis as the black dots, and indicate negative (unphysical) values of $r$ by the hatched dark grey region. 

We consider 3 dataset combinations: \textit{Planck} + BAO + lensing, BK18-only, and \textit{Planck} + BK18 + BAO + lensing. For each combination involving the \textit{Planck} data, we show the results obtained marginalising over the lowlEB covariance matrix (``SH'') and the ones conditioning it (``cond. HL''), as discussed in Section \ref{sec:data}. For each combination involving the BK18 data, we show the results fitting for the synchrotron spectral index $\beta_{\rm s}$ and the ones fixing it to $\beta_{\rm s}=-3.1$ (i.e. to the mean of the Gaussian prior imposed in the MCMC default analysis, see Subsection \ref{sec:priors}), labelled as ``fixed $\beta_{\rm s}$''. 

\begin{figure*}
    \centering
    \includegraphics[scale=0.59]{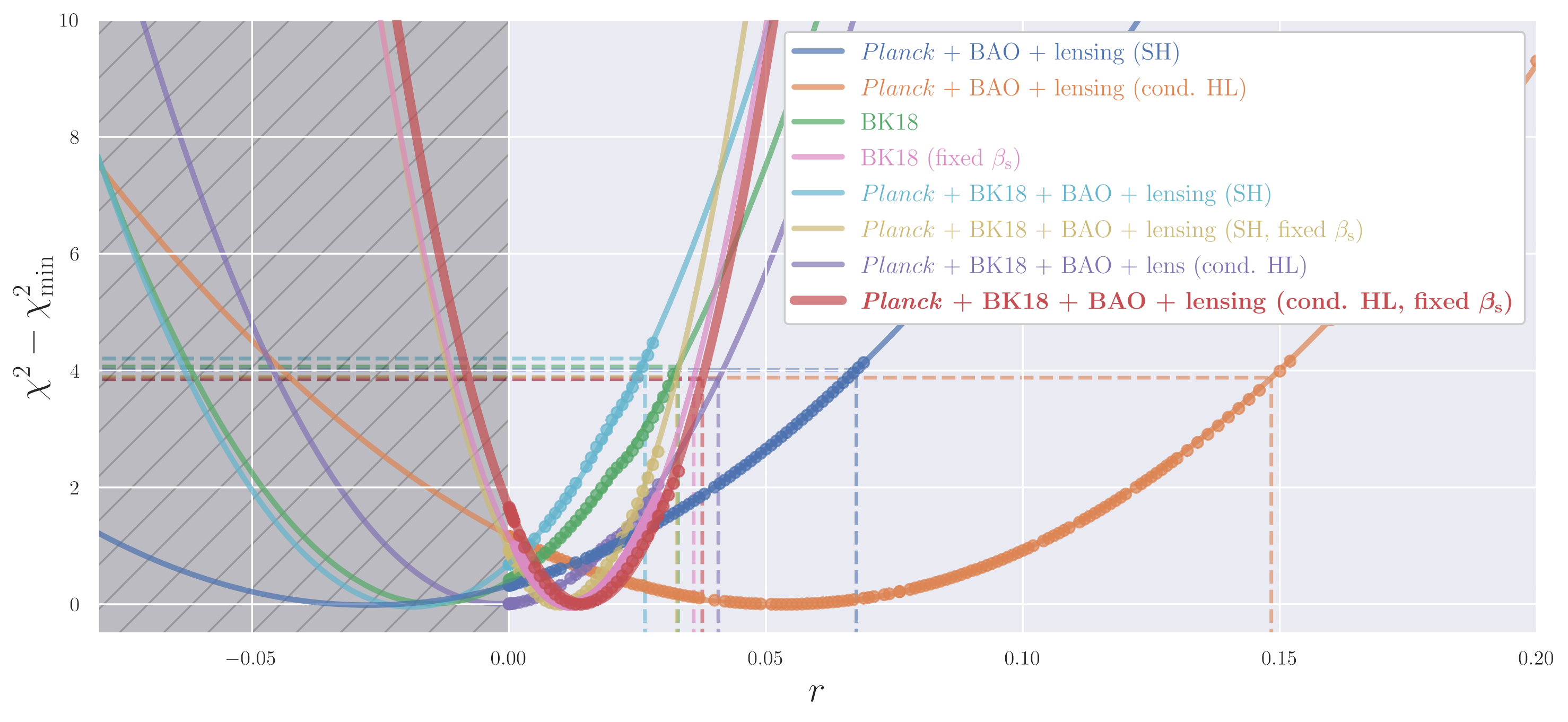}
    \caption{Profile likelihoods for $r$ from the datasets combinations considered in this work. The points are the $\chi^2 - \chi^2_{\rm min}$ values obtained from the likelihood maximization, whereas the parabolic fits are shown as the solid lines. The dashed lines indicate the upper limits at 95\% C.L. according to the FC prescription. Unphysical (negative) values of $r$ are shown as the dark grey hatched area. The baseline result is shown in the thick red line.}
    \label{fig:profile}
\end{figure*}

In Fig. \ref{fig:profile} we show $\Delta\chi^2(r)$ and the respective parabolic fits (solid lines), the upper limits from the FC prescription (vertical dashed lines), and the values of $\Delta\chi^2$ corresponding to each upper limit (horizontal dashed lines). The colours indicated in Table \ref{table:tab1} match those in Figures \ref{fig:barplot} and \ref{fig:profile}.      

We start by discussing the results of \textit{Planck} + BAO + lensing (blue and orange bars in Fig. \ref{fig:barplot}). If the SH correction is used, the MLE from the profile likelihood lies in the unphysical region of the parameter space ($r_{\rm MLE}=-0.027$), whereas conditioning the \textit{Planck} lowlEB covariance matrix shifts it to a large positive value ($r_{\rm MLE}=0.053$). This results in a significantly larger upper limit ($r<0.15$ instead of $r<0.068$) in the latter case, despite the width of $\Delta\chi^2(r)$ being the same in both cases. This confirms the findings of \citet{Beck:2022efr} in a prior-independent manner. 

Comparing the MCMC and profile likelihood limits, we observe that the limit from the profile likelihood is tighter in the SH-corrected case, whereas the opposite is true in the conditioned HL case. In the SH case, the MLE lies deep in the negative region, where Bayesian intervals notoriously overcover in the presence of a boundary \citep{Feldman:1997qc}. In the conditioned HL case, instead, the weaker profile likelihood limit is partly due to the different coverage properties and definitions of Bayesian limits compared to FC far from the boundary and partly to the effect of the Gaussian priors in the \textit{Planck} likelihood.  

The BK18-only constraints (green bars in Fig. \ref{fig:barplot}) are obtained fixing the $\Lambda$CDM parameters to their \citet{planck_2018} best-fitting values, as done in \citet{BICEP_2021}. 
The upper limit from the profile likelihood is slightly tighter than the corresponding MCMC case ($r<0.033$ versus $r<0.036$), with a MLE lying deep in the negative region. We notice, however, that the best-fitting model prefers a value $\beta_{\rm s} \simeq -2$ for the synchrotron spectral index nuisance parameter in the BK18 likelihood (see Appendix \ref{sec:appendix}). On the other hand, constraints on this parameter in the default MCMC analysis are prior-driven (see Section \ref{sec:priors}) and prefer a value $\beta_{\rm s} \simeq -3$. Therefore, a more straightforward comparison with the frequentist approach can be drawn after fixing $\beta_{\rm s}$ to the central value of the Gaussian prior in the profile likelihood (pink bars in Fig. \ref{fig:barplot}). We then recover $r<0.035$, very close to the Bayesian result given in \citet{BICEP_2021}.

   \begin{table*}
	\centering
	\begin{tabular}{ccccccc}
		\hline \hline
		\textbf{Data}
		& \textbf{Likelihood} 
		& \textbf{Fixed} $\mathbf{\beta_{\rm s}}$
		& \textbf{Profile (95\% C.L.)}
		& \textbf{MCMC (95\% C.L.)}
		& $\mathbf{r_{\rm \mathbf{MLE}}}$
		& \textbf{Colour}
		\\
		\hline	\hline
\multirow{2}{*}{\textit{Planck} + BAO + lensing} & SH & - & $r<0.068$ & $r<0.075$ & $-0.027$ & \crule[C0]{0.2cm}{0.2cm}  \\
& cond. HL & - & $r<0.15$ & $r<0.13$ & $0.053$ & \crule[C1]{0.2cm}{0.2cm}  \\ 
\hline
   \multirow{2}{*}{BK18} &  \multirow{2}{*}{fix $\Lambda$CDM params.} & \xmark  & $r<0.033$ & $r<0.036$ & $-0.015$ & \crule[C2]{0.2cm}{0.2cm}\\
    & & \cmark  & $r<0.035$ & $r<0.036$ & $0.013$ & \crule[C6]{0.2cm}{0.2cm}\\
\hline
\multirow{4}{*}{\textit{Planck} + BK18 + BAO + lensing}  & \multirow{2}{*}{SH} & \xmark  & $r<0.026$ & $r<0.032$ & $-0.019$ & \crule[C9]{0.2cm}{0.2cm}\\
 & & \cmark  & $r<0.032$  & $r<0.032$ & $0.01$ & \crule[C8]{0.2cm}{0.2cm}\\
  \cline{2-7}
 & \multirow{2}{*}{cond. HL} & \xmark & $r<0.042$ & $r<0.038$ & $-0.0021$ & \crule[C4]{0.2cm}{0.2cm}\\
 &  & \cmark  & $\mathbf{r<0.037}$  & $r<0.038$ & $0.015$  & \crule[C3]{0.2cm}{0.2cm}\\
 		\hline\hline
	\end{tabular}
	\caption{Upper limits on the tensor-to-scalar ratio parameter $r$ (95\% C.L.) from the profile likelihood method with the FC prescription and the MCMC. We also report the MLE for $r$ obtained from the profile likelihood method. For ``cond. HL'' we adopt the conditioning prescription defined in \citet{Beck:2022efr} for the \textit{Planck} lowlEB likelihood. For ``SH'' we marginalise the likelihood over the covariance matrix \citep{Sellentin:2015waz}. Note that for BK18-only data we fix all 6 $\Lambda$CDM parameters to the best-fitting values given in \citet{planck_2018}. For each case involving the BK18 likelihood, we indicate whether we are fixing the synchrotron spectral index to $\beta_{\rm s}=-3.1$ (see Section \ref{sec:data} for details).  The baseline result of this work is highlighted in bold. The colours shown in the rightmost column match those in Figures \ref{fig:barplot} and \ref{fig:profile}.}
	\label{table:tab1}
\end{table*}

The difference between the FC and Bayesian limits obtained from the \textit{Planck}+BK18+BAO+lensing data, both in the SH-corrected and the conditioned HL cases (the light blue and purple bars in Fig. \ref{fig:barplot}), is also due to the prior-dominated constraint on $\beta_{\rm s}$ in the MCMC analysis, and to its consequent effect on the position of the MLE and the different width of $\Delta\chi^2$ (compare e.g. the purple and red solid lines).
Fixing $\beta_{\rm s}$ in the profile likelihood leads to equal or slightly tighter limits than the MCMC ones (see the pink, yellow and red bars). Specifically, these small differences can be fully ascribed to the overcoverage of the Bayesian limit near the boundary (Section \ref{sec:profile}), since we found that fixing $\beta_{\rm s}$ in the MCMC analysis produces the same upper limit as imposing the Gaussian prior on it.

We also checked the effect of fixing the \textit{Planck} likelihood nuisance parameters to the mean values of their Gaussian prior \citep{Tristram:2021tvh, Tristram:2020wbi} in the profile likelihood analysis. For the \textit{Planck}+BK18 combinations (with fixed $\beta_{\rm s}$ and conditioned HL covariance) this leads to the the same upper limit and MLE as when fitting those nuisance parameters. In other words, the constraint is not prior-dominated and the nuisance parameters are constrained by the data.

We address the relevance of volume effects due to marginalization in the context of Bayesian inference for $r$. As evident from Fig. \ref{fig:barplot} and Table \ref{table:tab1}, no substantial difference exists between MCMC marginalized limits and the prior-independent FC ones, as long as prior-dominated nuisance parameters such as $\beta_{\rm s}$ are fixed in the profile likelihood analysis. This suggests that volume effects do not play a prominent role in the Bayesian constraints.
We note also that, because of the inefficiency of the Metropolis-Hastings algorithm in sampling near the boundary when a uniform positive prior is imposed on $r$, an apparent lower limit $r>0$, which is entirely caused by the prior-dominated posterior, appears \citep{Hergt:2021qlh}. This issue can be addressed for instance with the adoption of a logarithmic prior on $r$ (introducing however a dependence of the constraints on the choice of the prior lower edge) as well as with the profile likelihood approach we adopt in this paper.

\section{Conclusions}\label{sec:conclusions}
In this paper, we derived confidence intervals on $r$ from the state-of-the-art CMB datasets \textit{Planck} and BK18 via the frequentist profile likelihood method, and compared with the Bayesian MCMC procedure typically adopted in the literature. 
This is a useful robustness test for a potential future detection of $r$ or for putting robust upper limits on this parameter, checking simultaneously for the dependence on priors and the volume effect upon marginalization in the Bayesian constraints. The profile likelihood is not affected by the inefficiency of the MCMC sampling near the boundary when a uniform prior is imposed on $r\ge 0$.  

We confirmed that the profile likelihood method can provide upper limits comparable with the MCMC ones. Specifically, we reported an upper limit of $r<0.042$ at 95\% C.L. for the combination of \textit{Planck}, BK18, BAO and lensing with a conditioned \textit{Planck} lowlEB covariance matrix as suggested in \citet{Beck:2022efr}. This limit is slightly more conservative than the corresponding MCMC limit of $r<0.038$. We find that the Bayesian constraint is driven by the Gaussian prior adopted for the synchrotron spectral index $\beta_{\rm s}$ in the BK18 likelihood. Fixing this nuisance parameter to the central value of the prior, $\beta_{\rm s}=-3.1$, we obtained an upper limit of $r<0.037$ from the profile likelihood, slightly tighter than the  MCMC limit because of the well-known overcoverage of Bayesian intervals near the parameter boundary. 

We also confirmed the findings in \citet{Beck:2022efr} regarding the conditioning of the \textit{Planck} lowlEB covariance matrix: the additional scatter due to the limited number of simulations used in the covariance matrix construction moves the MLE of the lowlEB likelihood towards lower values, deceptively tightening the resulting upper limit.  

Given the consistency of the limits from the profile likelihood and the MCMC approach (provided that we fix $\beta_{\rm s}$), we confirmed that the Bayesian limits set in \citet{Tristram:2021tvh} and \citet{Beck:2022efr} are not significantly affected by volume effects arising during marginalization or by differences due to the choice of model parameterization (i.e. implicit priors). 

The profile likelihood method is computationally more efficient than the MCMC, providing a useful alternative for a fast and robust evaluation of confidence limits near the physical boundary of a parameter.

Although we do not find substantial differences with respect to the standard Bayesian approach using the current data, we anticipate that the profile likelihood will represent a useful sanity check for prior effects in future and increasingly sensitive surveys, such as the BICEP array \citep{Moncelsi:2020ppj}, the Simons Observatory \citep{SimonsObservatory:2018koc}, the LiteBIRD satellite \citep{LiteBIRD:2022cnt} and the CMB-S4 \citep{CMB-S4:2016ple} experiments. 

\appendix
\section{Best-fitting parameters}\label{sec:appendix}
In Table \ref{table:tab2} we compare the best-fitting parameters for the \textit{Planck}+BK18+BAO+lensing combination and conditioned lowlEB covariance matrix with and without fixing the synchrotron spectral index $\beta_{\rm s}$ in the BK18 likelihood. See Section \ref{sec:data} and references therein for details on likelihoods and parameters used here. 

   \begin{table*}
   \renewcommand{\arraystretch}{1.15}
	\centering\scalebox{0.9}{
	\begin{tabular}{cccc}
		\hline \hline
		\textbf{Parameter}
		& \textbf{Type}
		& \textbf{Best-fit (free $\mathbf{\beta_{\rm s}}$)}
		& \textbf{Best-fit (fixed $\mathbf{\beta_{\rm s}}$)}
		\\
		\hline	\hline
	\textbf{Cosmological parameters}	& & &  \\
$r$   &  profile MLE   & $-0.0021$ &    $0.015$        \\
		$\theta_\mathrm{MC}$   &  free   &  $0.104062$  &   $0.0104062$        \\
  $\log(10^{10} A_\mathrm{s})$   &  free  &  $3.054$  &  $3.054$             \\
  $n_\mathrm{s}$   & free    &  $0.966$ &   $0.967$                           \\
  $\Omega_\mathrm{b}h^2$   & free    &  $0.0223$   & $0.0223$                  \\
  $\Omega_\mathrm{c}h^2$   & free    &  $0.119$   & $0.119$                    \\
  $\tau$   &  free  &   $0.0603$    & $0.0604$                      \\
    $A_\mathrm{s}$   & derived    & $2.12\times 10^{-9}$ &    $2.12\times 10^{-9}$         \\
  $H_0$   &  derived  &  $67.43$ &      $67.43$        \\
  $\sigma_8$   & derived    & $0.813$ &    $0.813$         \\
\hline
	\textbf{BK18 nuisance parameters}	& & &  \\
$\beta_{\rm s}$   &  free or fixed & $-2.0$ (free) &      $-3.1$ (fixed)                   \\
  $A_{\rm d}$   &  free  &  $4.433$  &   $4.397$                     \\
  $A_{\mathrm{sync}}$   &  free &   $0.17$  &  $0.517$                     \\
  $\alpha_{\rm d}$   & free   &  $-0.641$ &     $-0.657$              \\
  $\alpha_{\rm s}$   & free    &  $-1.9\times 10^{-7}$ &   $-4.3\times 10^{-6}$       \\
  $\beta_{\rm d}$   &   free &  $1.500$   & $1.484$                 \\
  
  $\epsilon$   &  free & 0.03  &    -0.131          \\
  \hline
  	\textbf{\textit{Planck} nuisance parameters}	& & &  \\
  $A_{\rm pl}$  & free (\texttt{HiLLiPoP}, lowlTT, lensing)  &  $1.00189$ &     $1.00192$                       \\
  $c_0(\mathrm{100A})$   &  free (\texttt{HiLLiPoP}) &    $3.85\times 10^{-3}$  &     $3.86\times 10^{-3}$                     \\
  $c_1(\mathrm{100B})$   &  free (\texttt{HiLLiPoP}) &  $-1.0053\times 10^{-2}$ &    $-1.0053\times10^{-2}$                   \\
  $c_3(\mathrm{143B})$   &  free (\texttt{HiLLiPoP}) & $-1.0031\times 10^{-2}$  &   $-1.0026\times 10^{-2}$                   \\
  $c_4(\mathrm{217A})$   &  free (\texttt{HiLLiPoP}) &  $-1.0053\times 10^{-2}$  &  $-1.0053\times 10^{-2}$                   \\
  $c_5(\mathrm{217B})$   &  free (\texttt{HiLLiPoP}) &  $-4.419\times 10^{-3}$  &  $-4.417\times 10^{-3}$                    \\
  $A^\mathrm{PS}(100x100)$   &  free (\texttt{HiLLiPoP}) &  $2.620\times 10^{2}$   &  $2.619\times 10^{2}$             \\
  $A^\mathrm{PS}(100x143)$   &  free (\texttt{HiLLiPoP}) &  $1.245\times 10^{-2}$   & $1.244\times 10^{2}$              \\
  $A^\mathrm{PS}(100x217)$   & free (\texttt{HiLLiPoP})  &   $84.71$   &     $84.65$             \\
  $A^\mathrm{PS}(143x143)$   &  free (\texttt{HiLLiPoP}) & $53.09$ &    $53.05$              \\
  $A^\mathrm{PS}(143x217)$   &  free (\texttt{HiLLiPoP}) &  $37.70$ &   $37.67$              \\
  $A^\mathrm{PS}(217x217)$   &  free (\texttt{HiLLiPoP}) &  $74.39$  &  $74.41$              \\
  $A_\mathrm{dust}^\mathrm{100}$   & free (\texttt{HiLLiPoP})   &  $1.694\times 10^{-2}$ &    $1.688\times 10^{-2}$         \\
  $A_\mathrm{dust}^\mathrm{143}$   & free (\texttt{HiLLiPoP})    &  $3.966\times 10^{-2}$  &  $3.963\times 10^{-2}$         \\
  $A_\mathrm{dust}^\mathrm{217}$   & free (\texttt{HiLLiPoP})    &  $0.1322$  &  $0.1322$         \\
  $A_\mathrm{SZ}$   &  free (\texttt{HiLLiPoP}) & $1.050$ &    $1.048$                        \\
  $A_\mathrm{CIB}$   &  free (\texttt{HiLLiPoP}) & $1.056$ &    $1.055$                       \\
  $A_\mathrm{kSZ}$   &  free (\texttt{HiLLiPoP}) & $7.722\times 10^{-5}$  &   $2.723\times 10^{-5}$                       \\
  $A_\mathrm{SZxCIB}$   &  free (\texttt{HiLLiPoP}) &   $3.961\times 10^{-5}$  &     $3.582\times 10^{-6}$                     \\
  $c_2(143A)$   &  fixed (\texttt{HiLLiPoP}) &  $0.0$ &   $0.0$                            \\
  $A_\mathrm{radio}^\mathrm{PS}$   & fixed (\texttt{HiLLiPoP})  & $0.0$ &      $0.0$                    \\
  $A_\mathrm{dust}^\mathrm{PS}$   &  fixed (\texttt{HiLLiPoP})  & $0.0$ &     $0.0$                    \\
  $A_\mathrm{dust}^\mathrm{100T}$   & derived (\texttt{HiLLiPoP})   & $1.694\times 10^{-2}$ &    $1.688\times 10^{-2}$         \\
  $A_\mathrm{dust}^\mathrm{143T}$   &  derived (\texttt{HiLLiPoP})  & $3.966\times 10^{-2}$  &   $3.963\times 10^{-2}$         \\
  $A_\mathrm{dust}^\mathrm{217T}$   & derived (\texttt{HiLLiPoP})  &  $0.1322$  &  $0.1322$         \\
  $A_\mathrm{dust}^\mathrm{100P}$   &  derived (\texttt{HiLLiPoP})  &  $1.694\times 10^{-2}$  &  $1.688\times 10^{-2}$         \\
  $A_\mathrm{dust}^\mathrm{143P}$   &  derived (\texttt{HiLLiPoP}) &  $3.966\times 10^{-2}$  &  $3.963\times 10^{-2}$         \\
  $A_\mathrm{dust}^\mathrm{217P}$   &  derived (\texttt{HiLLiPoP})  &  $0.1322$  &  $0.1322$         \\
  \hline
    	\textbf{$\mathbf{\chi^2}$ values}	& & &  \\
  $\chi^2_\mathrm{BAO}$   &  -  & $17.87$  &   $17.84$         \\
  $\chi^2_{\rm BK18}$   &  -  & $534.70$ &     $536.20$       \\
  $\chi^2_{\rm lowlEB}$   & -   & $156.50$ &    $155.65$         \\
  $\chi^2_{\rm hillipop}$   &  -  &  $30346.11$   &  $30345.86$         \\
  $\chi^2_{\rm lowlTT}$   & -   & $22.90$ &    $23.36$        \\
  $\chi^2_{\rm lensing}$   &  -  & $8.71$ &    $8.72$         \\
  $\chi^2_{\rm tot}$   &  -  & $31086.786$ & $31087.63$    \\

		\hline\hline
	\end{tabular}}
	\caption{Best-fitting parameters obtained from the combination \textit{Planck} + BK18 + BAO + lensing with the conditioned HL covariance matrix, fitting (``free $\beta_{\rm s}$'' column) or fixing (``fixed $\beta_{\rm s}$'' column) $\beta_{\rm s}$ in the BK18 likelihood.}
	\label{table:tab2}
\end{table*}

\section*{Acknowledgements}

We thank Lukas Heinrich, Laura Herold and Matthieu Tristram for useful and stimulating discussion. 
We acknowledge the use of the \texttt{iMinuit}, \texttt{Cobaya}, \texttt{LoLLipop} and \texttt{HiLLiPoP} codes.
This work was supported in part by the Deutsche Forschungsgemeinschaft (DFG, German Research Foundation) under Germany's Excellence Strategy - EXC-2094 - 390783311.
The numerical analyses in this work have been supported by the Max Planck Computing and Data Facility (MPCDF) computer clusters \textit{Cobra}, \textit{Freya} and \textit{Raven}. 

\section*{Data Availability}

The data underlying this article will be shared on request to the corresponding author.



\bibliographystyle{mnras}
\bibliography{example} 

\begin{thebibliography}{}
\makeatletter
\relax
\def\mn@urlcharsother{\let\do\@makeother \do\$\do\&\do\#\do\^\do\_\do\%\do\~}
\def\mn@doi{\begingroup\mn@urlcharsother \@ifnextchar [ {\mn@doi@}
  {\mn@doi@[]}}
\def\mn@doi@[#1]#2{\def\@tempa{#1}\ifx\@tempa\@empty \href
  {http://dx.doi.org/#2} {doi:#2}\else \href {http://dx.doi.org/#2} {#1}\fi
  \endgroup}
\def\mn@eprint#1#2{\mn@eprint@#1:#2::\@nil}
\def\mn@eprint@arXiv#1{\href {http://arxiv.org/abs/#1} {{\tt arXiv:#1}}}
\def\mn@eprint@dblp#1{\href {http://dblp.uni-trier.de/rec/bibtex/#1.xml}
  {dblp:#1}}
\def\mn@eprint@#1:#2:#3:#4\@nil{\def\@tempa {#1}\def\@tempb {#2}\def\@tempc
  {#3}\ifx \@tempc \@empty \let \@tempc \@tempb \let \@tempb \@tempa \fi \ifx
  \@tempb \@empty \def\@tempb {arXiv}\fi \@ifundefined
  {mn@eprint@\@tempb}{\@tempb:\@tempc}{\expandafter \expandafter \csname
  mn@eprint@\@tempb\endcsname \expandafter{\@tempc}}}

\bibitem[\protect\citeauthoryear{{{ATLAS Collaboration}}}{{{ATLAS
  Collaboration}}}{2013}]{ATLAS:2013mma}
{{ATLAS Collaboration}} 2013, Report {ATLAS-CONF-2013-014}, {Combined
  measurements of the mass and signal strength of the Higgs-like boson with the
  ATLAS detector using up to 25 fb$^{-1}$ of proton-proton collision data}

\bibitem[\protect\citeauthoryear{Abazajian et~al.}{Abazajian
  et~al.}{2016}]{CMB-S4:2016ple}
Abazajian K.~N.,  et~al., 2016, {preprint} (\mn@eprint {arXiv} {1610.02743})

\bibitem[\protect\citeauthoryear{Ade et~al.}{Ade
  et~al.}{2019}]{SimonsObservatory:2018koc}
Ade P.,  et~al., 2019, \mn@doi [JCAP] {10.1088/1475-7516/2019/02/056}, 02, 056

\bibitem[\protect\citeauthoryear{Ade et~al.}{Ade et~al.}{2022}]{SPIDER:2021ncy}
Ade P. A.~R.,  et~al., 2022, \mn@doi [Astrophys. J.]
  {10.3847/1538-4357/ac20df}, 927, 174

\bibitem[\protect\citeauthoryear{Alam et~al.}{Alam
  et~al.}{2021}]{eBOSS:2020yzd}
Alam S.,  et~al., 2021, \mn@doi [Phys. Rev. D] {10.1103/PhysRevD.103.083533},
  103, 083533

\bibitem[\protect\citeauthoryear{Albrecht \& Steinhardt}{Albrecht \&
  Steinhardt}{1982}]{Albrecht:1982wi}
Albrecht A.,  Steinhardt P.~J.,  1982, \mn@doi [Phys. Rev. Lett.]
  {10.1103/PhysRevLett.48.1220}, 48, 1220

\bibitem[\protect\citeauthoryear{{{BICEP/Keck Collaboration}}}{{{BICEP/Keck
  Collaboration}}}{2021}]{BICEP_2021}
{{BICEP/Keck Collaboration}} 2021, \mn@doi [\prl]
  {10.1103/PhysRevLett.127.151301}, 127, 151301

\bibitem[\protect\citeauthoryear{Beck, Cukierman  \& Wu}{Beck
  et~al.}{2022}]{Beck:2022efr}
Beck D.,  Cukierman A.,   Wu W. L.~K.,  2022, {preprint} (\mn@eprint {arXiv}
  {2202.05949})

\bibitem[\protect\citeauthoryear{Campeti, Komatsu, Poletti  \&
  Baccigalupi}{Campeti et~al.}{2021}]{Campeti:2020xwn}
Campeti P.,  Komatsu E.,  Poletti D.,   Baccigalupi C.,  2021, \mn@doi [JCAP]
  {10.1088/1475-7516/2021/01/012}, 01, 012

\bibitem[\protect\citeauthoryear{Couchot, Henrot-Versill\'e, Perdereau,
  Plaszczynski, Rouill\'e~d'Orfeuil, Spinelli  \& Tristram}{Couchot
  et~al.}{2017}]{Couchot:2016vaq}
Couchot F.,  Henrot-Versill\'e S.,  Perdereau O.,  Plaszczynski S.,
  Rouill\'e~d'Orfeuil B.,  Spinelli M.,   Tristram M.,  2017, \mn@doi [Astron.
  Astrophys.] {10.1051/0004-6361/201629815}, 602, A41

\bibitem[\protect\citeauthoryear{Cousins}{Cousins}{1995}]{Cousins:1994yw}
Cousins R.~D.,  1995, \mn@doi [Am. J. Phys.] {10.1119/1.17901}, 63, 398

\bibitem[\protect\citeauthoryear{Cousins}{Cousins}{2018}]{Cousins:2018tiz}
Cousins R.~D.,  2018, {preprint} (\mn@eprint {arXiv} {1807.05996})

\bibitem[\protect\citeauthoryear{Feldman \& Cousins}{Feldman \&
  Cousins}{1998}]{Feldman:1997qc}
Feldman G.~J.,  Cousins R.~D.,  1998, \mn@doi [Phys. Rev. D]
  {10.1103/PhysRevD.57.3873}, 57, 3873

\bibitem[\protect\citeauthoryear{Fuskeland, Wehus, Eriksen  \&
  N\ae{}ss}{Fuskeland et~al.}{2014}]{Fuskeland:2014eoa}
Fuskeland U.,  Wehus I.~K.,  Eriksen H.~K.,   N\ae{}ss S.~K.,  2014, \mn@doi
  [Astrophys. J.] {10.1088/0004-637X/790/2/104}, 790, 104

\bibitem[\protect\citeauthoryear{G\'omez-Valent}{G\'omez-Valent}{2022}]{Gomez-Valent:2022hkb}
G\'omez-Valent A.,  2022, {preprint} (\mn@eprint {arXiv} {2203.16285})

\bibitem[\protect\citeauthoryear{Grishchuk}{Grishchuk}{1974}]{Grishchuk:1974ny}
Grishchuk L.~P.,  1974, Zh. Eksp. Teor. Fiz., 67, 825

\bibitem[\protect\citeauthoryear{Guth}{Guth}{1981}]{Guth:1980zm}
Guth A.~H.,  1981, \mn@doi [Phys. Rev. D] {10.1103/PhysRevD.23.347}, 23, 347

\bibitem[\protect\citeauthoryear{Hamann, Hannestad, Raffelt  \& Wong}{Hamann
  et~al.}{2007}]{Hamann:2007pi}
Hamann J.,  Hannestad S.,  Raffelt G.~G.,   Wong Y. Y.~Y.,  2007, \mn@doi
  [JCAP] {10.1088/1475-7516/2007/08/021}, 08, 021

\bibitem[\protect\citeauthoryear{Hamimeche \& Lewis}{Hamimeche \&
  Lewis}{2008}]{Hamimeche:2008ai}
Hamimeche S.,  Lewis A.,  2008, \mn@doi [Phys. Rev. D]
  {10.1103/PhysRevD.77.103013}, 77, 103013

\bibitem[\protect\citeauthoryear{Hergt, Handley, Hobson  \& Lasenby}{Hergt
  et~al.}{2021}]{Hergt:2021qlh}
Hergt L.~T.,  Handley W.~J.,  Hobson M.~P.,   Lasenby A.~N.,  2021, \mn@doi
  [Phys. Rev. D] {10.1103/PhysRevD.103.123511}, 103, 123511

\bibitem[\protect\citeauthoryear{Herold, Ferreira  \& Komatsu}{Herold
  et~al.}{2022}]{Herold:2021ksg}
Herold L.,  Ferreira E. G.~M.,   Komatsu E.,  2022, \mn@doi [Astrophys. J.
  Lett.] {10.3847/2041-8213/ac63a3}, 929, L16

\bibitem[\protect\citeauthoryear{Kamionkowski, Kosowsky  \&
  Stebbins}{Kamionkowski et~al.}{1997}]{Kamionkowski:1996zd}
Kamionkowski M.,  Kosowsky A.,   Stebbins A.,  1997, \mn@doi [Phys. Rev. Lett.]
  {10.1103/PhysRevLett.78.2058}, 78, 2058

\bibitem[\protect\citeauthoryear{{Lewis}, {Challinor}  \& {Lasenby}}{{Lewis}
  et~al.}{2000}]{camb}
{Lewis} A.,  {Challinor} A.,   {Lasenby} A.,  2000, \mn@doi [\apj]
  {10.1086/309179}, \href {http://adsabs.harvard.edu/abs/2000ApJ...538..473L}
  {538, 473}

\bibitem[\protect\citeauthoryear{Linde}{Linde}{1982}]{Linde:1981mu}
Linde A.~D.,  1982, \mn@doi [Phys. Lett. B] {10.1016/0370-2693(82)91219-9},
  108, 389

\bibitem[\protect\citeauthoryear{{{LiteBIRD Collaboration}}}{{{LiteBIRD
  Collaboration}}}{2022}]{LiteBIRD:2022cnt}
{{LiteBIRD Collaboration}} 2022, {preprint} (\mn@eprint {arXiv} {2202.02773})

\bibitem[\protect\citeauthoryear{Mangilli, Plaszczynski  \& Tristram}{Mangilli
  et~al.}{2015}]{Mangilli:2015xya}
Mangilli A.,  Plaszczynski S.,   Tristram M.,  2015, \mn@doi [Mon. Not. Roy.
  Astron. Soc.] {10.1093/mnras/stv1733}, 453, 3174

\bibitem[\protect\citeauthoryear{Moncelsi et~al.}{Moncelsi
  et~al.}{2020}]{Moncelsi:2020ppj}
Moncelsi L.,  et~al., 2020, \mn@doi [Proc. SPIE Int. Soc. Opt. Eng.]
  {10.1117/12.2561995}, 11453, 1145314

\bibitem[\protect\citeauthoryear{{{Planck Collaboration Int. XVI}}}{{{Planck
  Collaboration Int. XVI}}}{2014}]{planck_profile}
{{Planck Collaboration Int. XVI}} 2014, \mn@doi [Astron. Astrophys.]
  {10.1051/0004-6361/201323003}, 566, A54

\bibitem[\protect\citeauthoryear{{{Planck Collaboration V}}}{{{Planck
  Collaboration V}}}{2020}]{planck_2018_like}
{{Planck Collaboration V}} 2020, \mn@doi [Astron. Astrophys.]
  {10.1051/0004-6361/201936386}, 641, A5

\bibitem[\protect\citeauthoryear{{{Planck Collaboration VI}}}{{{Planck
  Collaboration VI}}}{2020}]{planck_2018}
{{Planck Collaboration VI}} 2020, \mn@doi [Astron. Astrophys.]
  {10.1051/0004-6361/201833910}, 641, A6

\bibitem[\protect\citeauthoryear{{{Planck Collaboration VIII}}}{{{Planck
  Collaboration VIII}}}{2020}]{Planck:2018lbu}
{{Planck Collaboration VIII}} 2020, \mn@doi [Astron. Astrophys.]
  {10.1051/0004-6361/201833886}, 641, A8

\bibitem[\protect\citeauthoryear{{{Planck Collaboration XI}}}{{{Planck
  Collaboration XI}}}{2016}]{planck_2015_like}
{{Planck Collaboration XI}} 2016, \mn@doi [Astron. Astrophys.]
  {10.1051/0004-6361/201526926}, 594, A11

\bibitem[\protect\citeauthoryear{{{Planck collaboration XV}}}{{{Planck
  collaboration XV}}}{2014}]{Planck:2013win}
{{Planck collaboration XV}} 2014, \mn@doi [Astron. Astrophys.]
  {10.1051/0004-6361/201321573}, 571, A15

\bibitem[\protect\citeauthoryear{Sato}{Sato}{1981}]{Sato:1980yn}
Sato K.,  1981, Mon. Not. Roy. Astron. Soc., 195, 467

\bibitem[\protect\citeauthoryear{Seljak \& Zaldarriaga}{Seljak \&
  Zaldarriaga}{1997}]{Seljak:1996gy}
Seljak U.,  Zaldarriaga M.,  1997, \mn@doi [Phys. Rev. Lett.]
  {10.1103/PhysRevLett.78.2054}, 78, 2054

\bibitem[\protect\citeauthoryear{Sellentin \& Heavens}{Sellentin \&
  Heavens}{2016}]{Sellentin:2015waz}
Sellentin E.,  Heavens A.~F.,  2016, \mn@doi [Mon. Not. Roy. Astron. Soc.]
  {10.1093/mnrasl/slv190}, 456, L132

\bibitem[\protect\citeauthoryear{Starobinsky}{Starobinsky}{1979}]{Starobinsky:1979ty}
Starobinsky A.~A.,  1979, JETP Lett., 30, 682

\bibitem[\protect\citeauthoryear{Torrado \& Lewis}{Torrado \&
  Lewis}{2021}]{Torrado:2020dgo}
Torrado J.,  Lewis A.,  2021, \mn@doi [JCAP] {10.1088/1475-7516/2021/05/057},
  05, 057

\bibitem[\protect\citeauthoryear{Tristram et~al.}{Tristram
  et~al.}{2021}]{Tristram:2020wbi}
Tristram M.,  et~al., 2021, \mn@doi [Astron. Astrophys.]
  {10.1051/0004-6361/202039585}, 647, A128

\bibitem[\protect\citeauthoryear{Tristram et~al.}{Tristram
  et~al.}{2022}]{Tristram:2021tvh}
Tristram M.,  et~al., 2022, \mn@doi [Phys. Rev. D]
  {10.1103/PhysRevD.105.083524}, 105, 083524

\bibitem[\protect\citeauthoryear{Trotta}{Trotta}{2017}]{Trotta:2017wnx}
Trotta R.,  2017, {preprint} (\mn@eprint {arXiv} {1701.01467})

\bibitem[\protect\citeauthoryear{{Virtanen} et~al.,}{{Virtanen}
  et~al.}{2020}]{scipy}
{Virtanen} P.,  et~al., 2020, \mn@doi [Nature Methods]
  {10.1038/s41592-019-0686-2}, \href
  {https://ui.adsabs.harvard.edu/abs/2020NatMe..17..261V} {17, 261}

\bibitem[\protect\citeauthoryear{Zyla et~al.}{Zyla
  et~al.}{2020}]{ParticleDataGroup:2020ssz}
Zyla P.~A.,  et~al., 2020, \mn@doi [PTEP] {10.1093/ptep/ptaa104}, 2020, 083C01

\makeatother
\end{thebibliography}




\appendix


\bsp	
\label{lastpage}
\end{document}